\documentclass[12pt]{article}

\usepackage{epsfig,amsmath,amssymb,graphics,graphicx} 

\newcommand{\be}{\begin{equation}}
\newcommand{\ee}{\end{equation}}
\newcommand{\bi}[1]{\vspace{-3mm} \bibitem{#1}}

\begin{document}

\today

\begin{center}
{\Large \bf 
Coupled oscillators with power-law interaction and their fractional dynamics analogues}
\vskip 5 mm

{\large \bf Nickolay Korabel$^{a,}$\footnote{Corresponding author. Tel.: +1 212 998 3260; fax: +1 212 995 4640. \\ E-mail address: korabel@cims.nyu.edu.}, George M. Zaslavsky$^{a,b}$ \\ and Vasily E. Tarasov$^{c}$ } \\

\vskip 3mm

{\it $^a$ Courant Institute of Mathematical Sciences, New York University \\
251 Mercer Street, New York, NY 10012, USA, }\\
{\it $^b$ Department of Physics, New York University, \\
2-4 Washington Place, New York, NY 10003, USA } \\ 
{\it $^c$ Skobeltsyn Institute of Nuclear Physics, \\
Moscow State University, Moscow 119992, Russia } \\
\end{center}

\vskip 11 mm

\begin{abstract}
The one-dimensional chain of coupled 
oscillators with long-range power-law interaction is considered.
The equation of motion in the infrared limit 
are mapped onto the continuum equation
with the Riesz fractional derivative of order $\alpha$, 
when $0<\alpha<2$. The evolution of soliton-like and breather-like 
structures are obtained numerically and compared for both types of simulations: 
using the chain of oscillators and using the continuous medium 
equation with the fractional derivative. 
\end{abstract}

\vskip 3 mm
{\small 

\noindent
{\it PACS}: 45.05.+x; 45.50.-j; 45.10.Hj


\vskip 3 mm

\noindent
{\it Keywords}: Long-range interaction, Fractional oscillator, 
Fractional equations

\vskip 11 mm

\section{Introduction}

Nowadays it become clear that anomalous dynamics and 
kinetics may be not so pathological as it was believed earlier. 
More and more situations has been reported in the 
literature supporting this \cite{BG,Zaslavsky1,Montr,SZK,Hilfer}.  
The immediate causes for these anomalies could be attributed to 
fractal or multi-fractal character of phase space, L{\`e}vy flights, 
dynamical traps, or long-range correlations which are present 
in many interesting applications \cite{Zaslavsky1,Montr,SZ,Uch}.  
The desolated once fractional order calculus revives again in order to discribe such problems. 
Indeed, equations which involve derivatives or integrals of non-integer order 
appeared to be very successful in describing anomalous 
kinetics and transport \cite{SZ,SZK,KSZ,SKB,MK,KS,MS2,Hilfer}. 

However, fractional equations could be rarely derived explicitly from the equations 
of motion or from a Hamiltonian of the model. 
More often fractional equations for dynamics or kinetics 
appear as some phenomenological models. 
Recently, the method to obtain fractional analogues of equations 
of motion was considered for the 
bunch of problems related to sets of coupled particles 
or other objects that interact via a power-like potential \cite{LZ,TZ_a}. 
Examples of such systems are one-dimensional chain of interacting oscillators, spins, or waves that
can be considered as a benchmark for numerous applications 
in physics, chemistry, biology, etc. \cite{Dys,CMP,Kur,Ruf,BrKiv,Br4,PV}. Transfer from 
the set of Hamiltonian equations to the continuous media equation 
with fractional derivatives is an approximate procedure. 
Its formal realization can be perfomed  using the so-called 
"transform operator" \cite{TZ_a}. 
Different applications of the operator have already been used 
to derive fractional sine-Gordon and fractional 
Hilbert equation \cite{LZ}, to study synchronization of 
coupled oscillators \cite{TZ_a}, and for fractional 
Ginzburg-Landau equation \cite{TZ}.

Localized in space structures, such as solitons or breathers are the subjects 
of a great interest since a long time. 
These structures have been widely studied in discrete systems on lattices 
with different types of interactions as well as 
in their continuous analogues (for a review, see \cite{FW,BK}).  
Solitons in a one-dimensional lattice with the 
long-range Lennard-Jones-type interaction were considered in \cite{Ish}. 
Kinks in the Frenkel-Kontorova model with long-range interparticle interactions 
were studied in \cite{BKZ}. The properties of time periodic spatially localized solutions (breathers) 
on discrete chains in the presence of algebraically decaying interactions 
were considered in \cite{Br4} and recently in \cite{GF}. 
Energy and decay properties of discrete breathers in systems with long-range 
interactions have also been studied in the framework of the Klein-Gordon 
\cite{BK}, and discrete nonlinear Schrodinger equations \cite{Br6}. 
A remarkable property of the dynamics described by the equation with 
fractional space derivatives is that the soliton or breather 
type solutions have power-like tails. Similar features were observed 
in the prototype lattice models with power-like long-range 
interactions \cite{Br4,GF,PV,AEL,AK}. As it was shown in \cite{TZ_a,TZ}, 
analysis of the equations with fractional derivatives (FD) can provide 
fairly quickly results for the space asymptotics of their solutions. 
Replacement of the system of particles with power long-range interactions 
by an equation with FD will be called fractional equation approximation (FEA). 
As it was shown in \cite{TZ_a,TZ}, the FEA appears in the infrared limit when the 
wave number $k \rightarrow 0$ and some other conditions that will be derived later. 
At the same time, in many applied problems equations with FD can appear as an 
intrinsic feature of the system (see examples in reviews \cite{Zaslavsky1,MK}). 
Integration of equations with FD needs specific algorithms which are fairly time 
consuming with not yet well defined errors \cite{MT}.

The goal of this paper is to study a duality of Hamiltonian 
dynamics of system of particles with power-like interactions with 
the solutions of equations with FD using the transform operator. 
As a model to be studied we consider a chain of interacting oscillators 
that can be described by the sine-Gordon equation in the continuous limit.

To incorporate long-range interparticle interactions several 
nonlocal generalizations of the sine-Gordon equation were considered in Refs. \cite{PV,AEL,AK} 
using the integro-differential equations of motion.
The generalization of the standard sine-Gordon equation 
with space FD was proposed and 
numerically studied in \cite{APV}. 
The form of the fractional Sine-Gordon equation was 
$u_{tt} - {}^{R} D^{\alpha} u + \sin u = 0$, 
where ${}^{R} D^{\alpha}$ is the Riesz space fractional derivative, $1 < \alpha < 2$.  
This equation was proposed as an interpolation between the classical sine-Gordon equation 
and a nonlocal sine-Gordon equation. Here, we consider the one-dimentional lattice 
of coupled non-linear oscillators. We focus on the 
dynamics of soliton-like and breather-like patterns. 
Following \cite{LZ}, we make the transform to the infrared limit and derive 
the fractional sine-Gordon equation which describes the dynamics of the lattice.
Using both, Hamiltonian lattice dynamics and the fractional 
sine-Gordon (FSD) equation, we can compare solutions and 
demonstrate their similarity as well as to discuss conditions when such 
duality is applicable.

The obtained results can be used in a twofold way: 
firstly, we show how the infrared limit for the chain 
of oscillators can be described by the corresponding 
FSG equation with similar solutions; secondly, 
we propose the replacement of the integration scheme 
of equations with FD by the system of coupled Hamiltonian equations. 
The latter can be considered as a new way of analysis of the equations with FD.


\section{Long-range interaction of oscillators}

Consider a one-dimensional system of interacting oscillators described by the Hamiltonian
\be \label{H}
H =  \sum_{n=-\infty}^{+\infty} \left[ \frac{M}{2} \; 
\dot{u}_n^2 + \frac{J_0}{2} \sum_{\substack{m=-\infty \\ n \ne m}}^{+\infty} \; 
\frac{1}{|n-m|^{1+\alpha}} \; u_n u_m + \frac{J_1}{2} \; 
u_n^2 + J_2 \left( 1 - \cos \left(\frac{2 \pi u_n}{a}\right) \right) \right],
\ee
where $M$ is mass and $u_n$ is displacement from the equilibrium. 
The last two terms characterize an interaction of the chain 
with the external on-site potential which is defined 
by the periodic function with period $a$ and amplitude $J_2$. 
The second term in Eq.\ (\ref{H}) takes into account the interaction 
of the oscillators in the chain. Here, we consider nonlocal coupling 
given by the power-law function. Constant $\alpha$ is a physical relevant parameter.
Some integer values of $\alpha$ corresponds to the well-known physical situations: 
Coulomb potential corresponds to $\alpha=0$, dipole-dipole interaction 
corresponds to $\alpha=2$, and the limit $\alpha \rightarrow \infty$ 
is for the case of nearest-neighbor interaction.

From Hamiltonian (\ref{H}) it follows the equation of motion,
\be \label{Main_Eq}
\frac{\partial^2 u_n}{\partial t^2} + J_0 
\sum_{\substack{m=-\infty \\ n \ne m}}^{+\infty} \; 
\frac{1}{|n-m|^{1+\alpha}} \; u_m + J_1 u_n + J_2 \sin \left( u_n \right) = 0,
\ee
where we put $M=1$ and the period $a=2 \pi$. 

Let us sketch the derivation of continuous medium equation. We define the field $\hat{u}(k,t)$ on $[-K/2, K/2]$ as 
\be \label{ukt}
\hat{u}(k,t) = \sum_{n=-\infty}^{+\infty} \; u_n(t) \; e^{-i k x_n} = {\cal F}_{\Delta} \{u_n(t)\} ,
\ee
where  $x_n = n \Delta x$, and $\Delta x=2\pi/K$ is distance between oscillators
\be \label{un} 
u_n(t) = \frac{1}{K} \int_{-K/2}^{+K/2} dk \ \hat{u}(k,t) \; e^{i k x_n}= 
{\cal F}^{-1}_{\Delta} \{ \hat{u}(k,t) \} . 
\ee
These equations are the basis for the Fourier transform which is obtained by transforming $u_n(t)=u(n \Delta x,t)$ from discrete variable $x_n=n \Delta x$ to a continuous one in the limit as $\Delta x \rightarrow 0$ ($K \rightarrow \infty$). The Fourier transform is a generalization of (\ref{ukt}), (\ref{un}) in the limit as $\Delta x \rightarrow 0$.
Replace the discrete $u_n(t)$ with continuous $u(x,t)$ 
while letting $x_n=n\Delta x= 2\pi n/K \rightarrow x$.
Then change the sum to an integral, and 
Eqs. (\ref{ukt}), (\ref{un}) become
\be \label{ukt2} 
\tilde{u}(k,t)=\int^{+\infty}_{-\infty} dx \ e^{-ikx} u(x,t) = {\cal F} \{ u(x,t) \}, 
\ee
\be \label{uxt}
u(x,t)=\frac{1}{2\pi} \int^{+\infty}_{-\infty} dk \ e^{ikx} \tilde{u}(k,t) = {\cal F}^{-1} \{ \tilde{u}(k,t) \}, 
\ee
where
\be 
\tilde{u}(k,t)= {\cal L} \hat{u}(k,t), \quad u(x,t)= {\cal L} u_n(t), \quad u_n(t)=u(x_n,t), 
\ee
and ${\cal L}$ denotes the passage to the limit $\Delta x \rightarrow 0$ ($K \rightarrow \infty$).
The procedure of the replacement of a discrete model 
by the continuous one is defined by the following operations:\\
1) The Fourier series transform:
\be \label{O1}
{\cal F}_{\Delta}: \quad u_n(t) \rightarrow {\cal F}_{\Delta}\{ u_n(t)\}=\hat{u}(k,t) ;
\ee
2) The passage to the limit $\Delta x \rightarrow 0$:
\be
{\cal L}: \quad \hat{u}(k,t) \rightarrow {\cal L} \{\hat{u}(k,t)\}=\tilde{u}(k,t) ;
\ee
3) The inverse Fourier transform:
\be
{\cal F}^{-1}: \quad \tilde{u}(k,t) \rightarrow 
{\cal F}^{-1} \{ \tilde{u}(k,t)\}=u(x,t) .
\ee
The operation $T={\cal F}^{-1} {\cal L} \ {\cal F}_{\Delta}$ 
can be called a transform operation (transform map), since it
performs a tranfrormation of a discrete model of coupled oscillators
by the continuous medium model. In a similar to \cite{LZ,TZ_a} way, we can obtain from (\ref{Main_Eq}) 
the equation for $\hat{u}(k,t)$ using (\ref{ukt}) 
\be \label{C3}
\frac{\partial^2 \hat{u}(k,t)}{\partial t^2} + J_0 \; 
\hat{J}_{\alpha}(k) \; \hat{u}(k,t) + J_1 \; \hat{u}(k,t) + 
J_2 \; \mathcal{F}_{\Delta} \{\sin \left( u_n(t) \right)\} = 0,
\ee
where 
\be \label{C5}
\hat{J}_{\alpha}(k) = \sum_{\substack{n=-\infty \\ n \ne 0}}^{+\infty} 
e^{-ikn\Delta x} \frac{1}{|n|^{1+\alpha}}= 
\sum^{+\infty}_{n=1} \frac{1}{n^{1+\alpha}} \left( e^{-ikn\Delta x} +e^{ikn\Delta x} \right) = 
Li_{1+\alpha}( e^{ik\Delta x} ) + Li_{1+\alpha}( e^{-ik\Delta x} ),
\ee
and $Li_{1+\alpha}(z)$ is a polylogarithm function. Using the series representation of the polylogarithm \cite{Erd}
\be \label{D1}
Li_{\beta}(e^z)=\Gamma(1-\beta) (-z)^{\beta-1}+\sum^{\infty}_{n=0}
\frac{\zeta(\beta-n)}{n!} z^n, \quad |z|< 2\pi, \; \; \beta\not=1,2,3...,
\ee
we obtain
\be \label{D2}
\hat{J}_{\alpha}(k)= a_{\alpha} \; |\Delta x|^{\alpha} \; |k|^{\alpha} +
2 \sum^{\infty}_{n=0} \frac{\zeta(1+\alpha-2n)}{(2n)!} (\Delta x)^{2n} (-k^2)^n , 
\quad |k|< 1, \quad \alpha \not=0,1,2,3...,
\ee
where $J_{\alpha}(0)=2 \zeta(1+\alpha)$, $\zeta$ is the Riemann zeta-function and
\be \label{D5}
a_{\alpha} =  2 \; \Gamma(-\alpha) \; \cos \left( \frac{\pi \alpha}{2} \right).
\ee

Function $\hat{J}_{\alpha}(k)$ can also be presented in the form
\be
\hat{J}_{\alpha}(k)=2\sum^{\infty}_{n=1} \frac{\cos(kn\Delta x)}{n^{1+\alpha}}, 
\ee
from which it follows that
\be
\hat{J}_{\alpha}(k+2\pi m/\Delta x)= \hat{J}_{\alpha}(k) ,
\ee
where $m$ is an integer. 
For $\alpha=2$, $\hat{J}_{\alpha}(k)$ is the Clausen function $Cl_{2}(k)$ \cite{Le}.

Combinig all expressions, Eq. (\ref{C3}) takes the form
\be \label{D4}
\frac{\partial^2 \hat{u}(k,t)}{\partial t^2} + J_0 \; a_{\alpha} |\Delta x|^{\alpha} \; |k|^{\alpha} \; 
\hat{u}(k,t) + 2 J_0 \sum^{\infty}_{n=0} \frac{\zeta(\alpha+1-2n)}{(2n)!} 
(\Delta x)^{2n} (-k^2)^n \hat{u}(k,t) +
\ee
\[+ J_1 \; \hat{u}(k,t) + J_2 \; \mathcal{F}_{\Delta} \{ \sin \left( u_n(t) \right) \} = 0.\]
We will be interested in the limit $\Delta k \rightarrow 0$. Then the Eq.\ (\ref{D4}) can be written in a simplified way
\be \label{Appr}
\frac{\partial^2}{\partial t^2} \hat{u}(k,t) + \bar{J}_0 \; \hat{\mathcal{T}}_{\alpha, \Delta}(k) \; \hat{u}(k,t) + J_1 \; \hat{u}(k,t) + J_2 \; \mathcal{F}_{\Delta} \{ \sin \left( u_n(t) \right) \} = 0, \quad 
\alpha \not=0,1,2,...,
\ee
where $\bar{J}_0=J_0 |\Delta x|^{min\{\alpha,2\}}$ and
\be \label{D8}
\hat{\mathcal{T}}_{\alpha, \Delta}(k) = \begin{cases} 
a_{\alpha} |k|^{\alpha} - |\Delta x|^{2-\alpha} \zeta (\alpha -1) k^2, & 0 < \alpha < 2, \quad (\alpha \not=1) 
\cr  
|\Delta x|^{\alpha-2} a_{\alpha} |k|^{\alpha} - \zeta (\alpha -1) k^2, & 2< \alpha < 4, \quad (\alpha \not=3).
\end{cases}
\ee
The expression (\ref{D8}) was obtained in \cite{LZ,TZ_a} in a slightly different way. 
Let us note that (\ref{D8}) has a crossover scale for 
\be \label{D10}
k_0 = |a_{\alpha}/\zeta (\alpha-1)|^{1/(2-\alpha)} |\Delta x|^{-1}
\ee
such that $\hat{\mathcal{T}}_{\alpha, \Delta}(k) \sim k^2$ for $\alpha>2$, $k \ll k_0$ 
and nontrivial expression 
$\hat{\mathcal{T}}_{\alpha, \Delta} (k) \sim |k|^{\alpha}$ appears only for $\alpha<2$, $k \ll k_0$. The crossover was considered also in \cite{TZ_a,Zaslavsky6}. The expression for $\hat{\mathcal{T}}_{\alpha,\Delta} (k)$ can be considered
as a Fourier transform of the operator 
\be \label{Z3}
\mathcal{T}_{\alpha, n} \; u_n(t) \equiv \sum_{\substack{m=-\infty \\ m \ne n}}^{+\infty} \; |n-m|^{-(1+\alpha)} \; u_m (t). 
\ee 
Performing the transition to the limit $k \ll k_0 $ (or more precisely $k \Delta x \ll k_0 \Delta x$), and applying inverse Fourier transform to (\ref{Appr}) gives
\be \label{D10b}
\frac{\partial^2}{\partial t^2} u(x,t) + \bar{J}_0 \; \mathcal{T}_{\alpha}(x) \; u(x,t) +
J_1 \; u(x,t) + J_2 \; \sin \left( u(x,t) \right) = 0 \quad \alpha \not=0,1,2,...,
\ee
where 
\be \label{Tx} 
\mathcal{T}_{\alpha}(x) = 
\mathcal{F}^{-1} \{ \hat{\mathcal{T}}_{\alpha} (k) \} = 
\begin{cases} 
- a_{\alpha} \frac{\partial^{\alpha}}{\partial |x|^{\alpha}} , & 0 < \alpha < 2, \quad (\alpha \not=1) \cr 
\zeta (\alpha -1) \frac{\partial^2}{\partial |x|^2}, & 2< \alpha < 4, \quad (\alpha \not=3);
\end{cases}
\ee
\[
\hat{\mathcal{T}}_{\alpha} (k) = 
\begin{cases} 
a_{\alpha} |k|^{\alpha}, & 0 < \alpha < 2, \quad (\alpha \not=1) \cr 
- \zeta (\alpha -1) \; k^2, & 2< \alpha < 4, \quad (\alpha \not=3).
\end{cases}
\]
Here, we have used the connection between the Riesz fractional derivative and its Fourier transform \cite{SKM}: 
\be
|k|^{\alpha} \longleftrightarrow - \frac{\partial^{\alpha}}{\partial |x|^{\alpha}}, \quad k^2 \longleftrightarrow - \frac{\partial^2}{\partial |x|^2}.
\ee
The properties of the Riesz derivative can be found in \cite{SKM,OS,MR,Podlubny}. 

If we consider a system with Hamiltonian $H=H_1+H_{int}$, 
where
\be \label{Z1} H_1 =  \sum_{n=-\infty}^{+\infty} \left[ 
\frac{1}{2} \; \dot{u}_n^2 + \; V (u_n)  \right], \quad
H_{int}= \frac{1}{2} J_0 
\sum_{\substack{n,m=-\infty \\ n \ne m}}^{+\infty} 
\frac{1}{|n-m|^{1+\alpha}} \; g(u_n) \; g(u_m) ,
\ee
where $g(u)$ is some function of $u$. 
Equation (\ref{H}) appears for $g(u)=u$.
The corresponding generalization of (\ref{D10b}) is easily obtained as 
\be
\frac{\partial^2 u(x,t)}{\partial t^2} = V^{\prime}(u(x,t))+
\bar{J}_0 \; g^{\prime}(u(x,t)) \mathcal{T}_{\alpha}(x) \; g(u(x,t)), 
\; \; \; (\alpha\not=0,1,2,...),
\ee
where $g^{\prime}(u)=\partial g(u)/\partial u$ and 
$\hat{ \mathcal{T}_{\alpha}}(x)$ is the same as in (\ref{Tx}).


\section{Numerical methods}

\begin{figure}
\centering
\rotatebox{0}{\includegraphics[width=7.9 cm,height=7.9 cm]{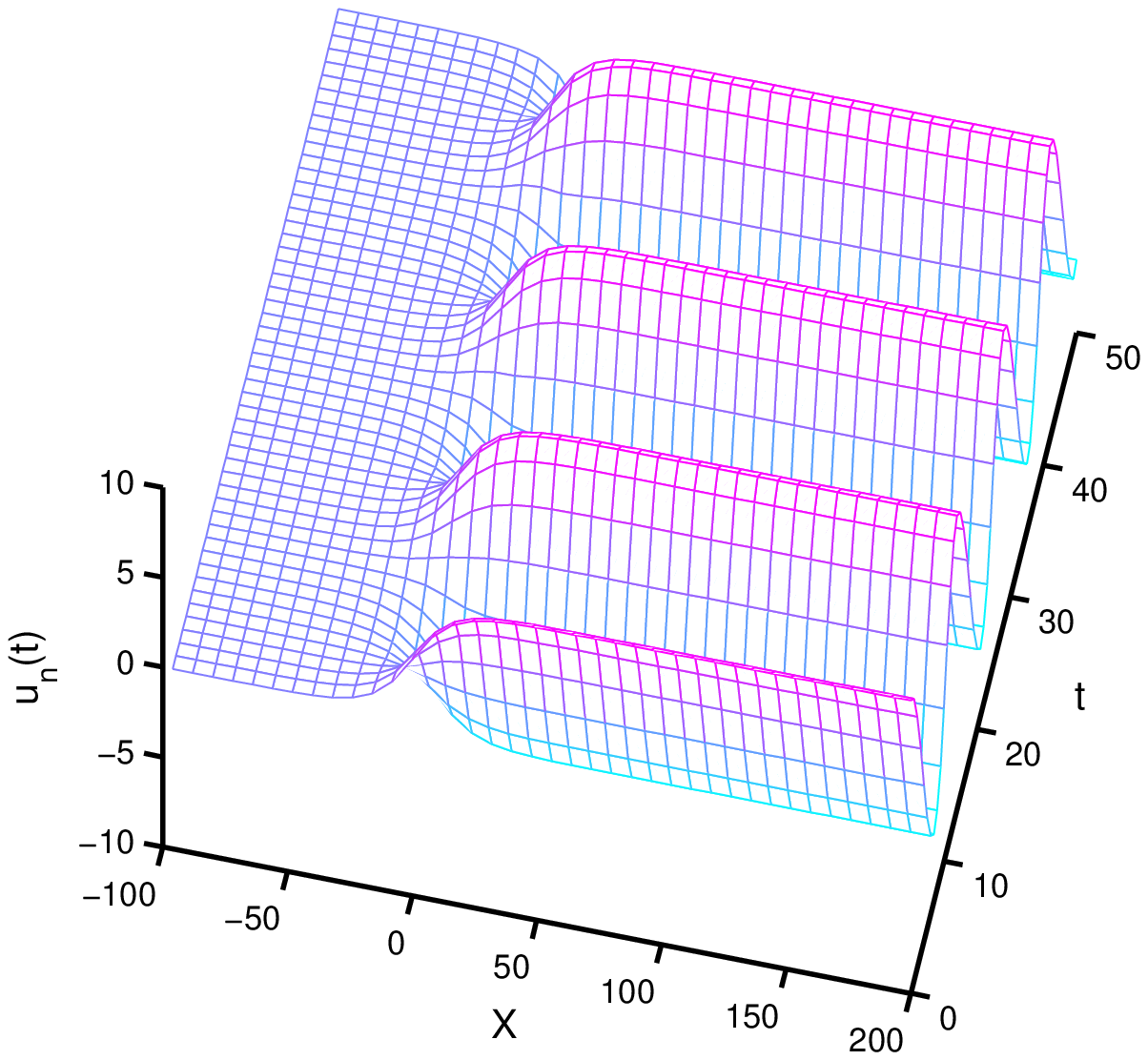}}
\rotatebox{0}{\includegraphics[width=7.9 cm,height=7.9 cm]{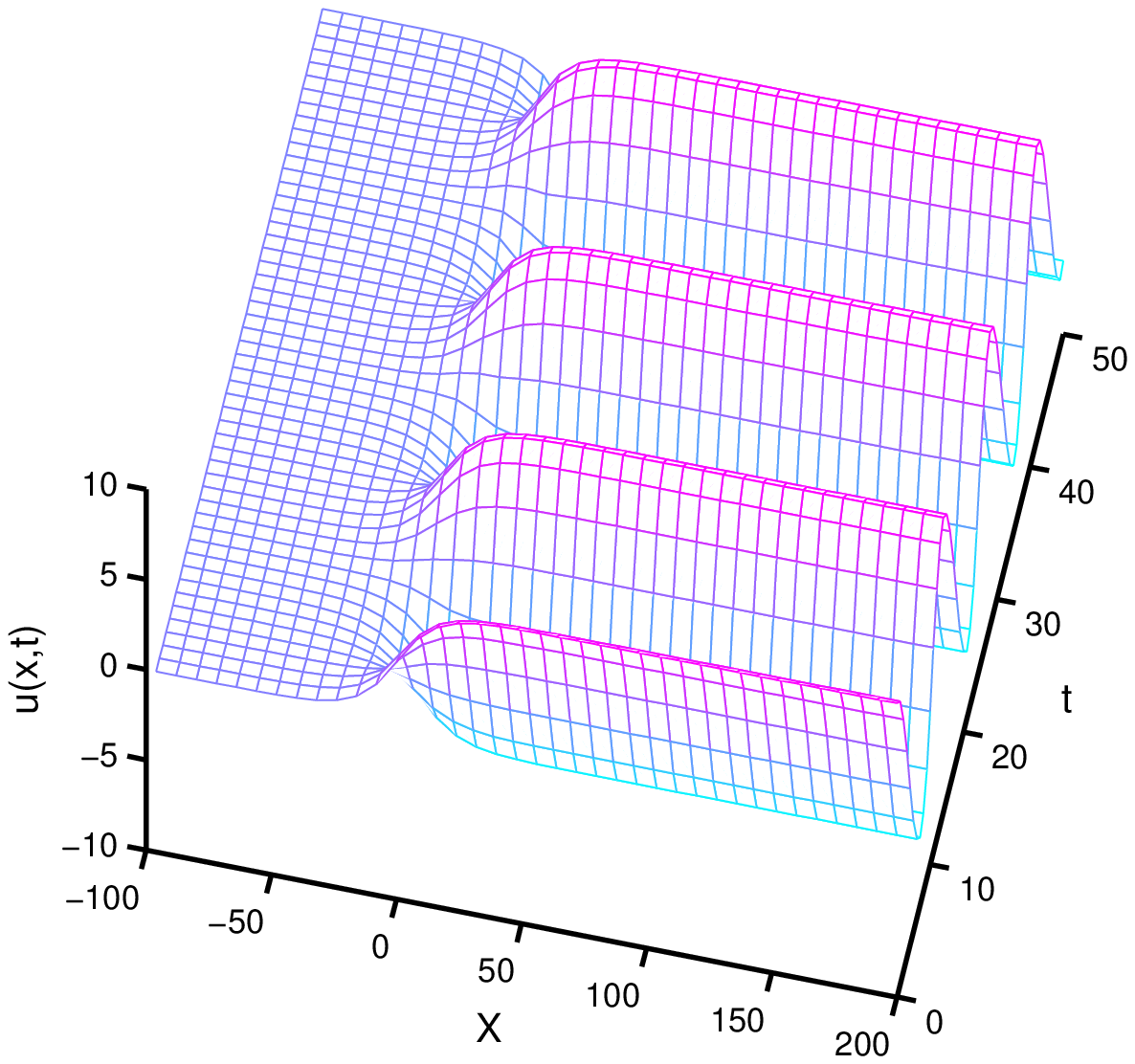}}
\rotatebox{0}{\includegraphics[width=7.8 cm,height=7.8 cm]{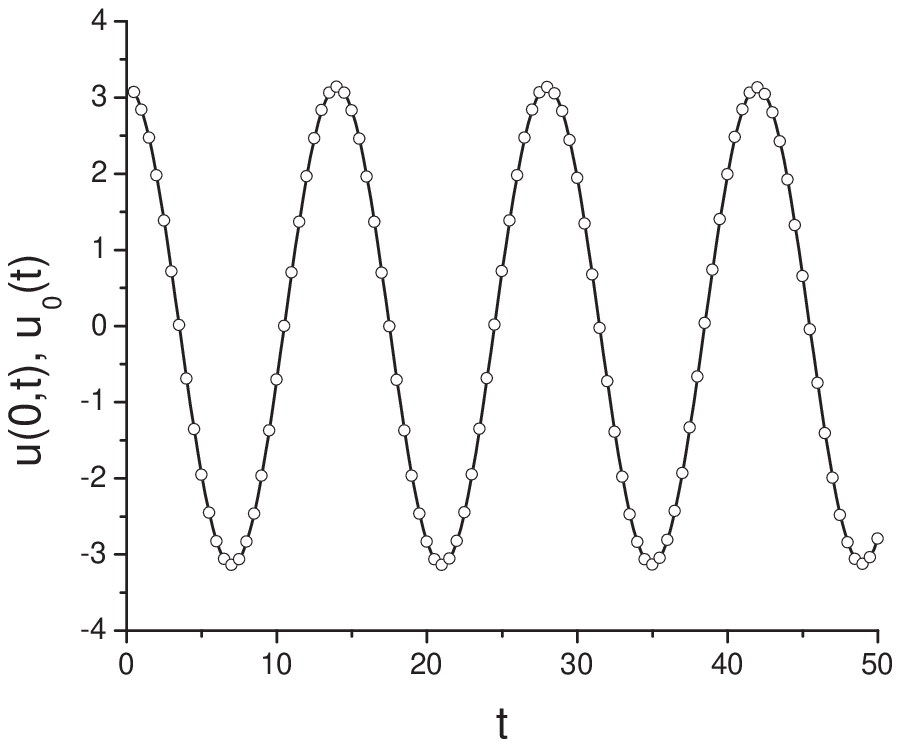}}
\caption{\label{fig1} 
Kink-like structures obtained from the evolution of the 
system of coupled oscillators (left panel) and from the 
solution of the FSG equation (right panel). 
Bottom panel: Comparison of the time evolution of the central 
oscillator $u_0(t)$ (solid curve) with  the time evolution of the $u(0,t)$ (circles) obtained 
from the solution of the FSG equation.
Exponent $\alpha=1.21$.
}
\end{figure}
 
We consider $N+1$ equally spaced by $\Delta x=2L/(N+1)$ coupled oscillators on the 
finite interval $(-L,L)$. The system is described by the equations of motion 
\be \label{Main_Eq2}
\frac{\partial^2 u_n}{\partial t^2} + J_0 \sum_{\substack{m=-N/2 \\ m \ne n}}^{N/2} \; 
\frac{1}{|n-m|^{1+\alpha}} \; u_m + J_1 u_n + J_2 \sin u_n = 0, \; \; 
(n=-\frac{N}{2}, ..., 0, ..., \frac{N}{2})
\ee
and $N$ is even. For $N \rightarrow \infty$ Eq.\ (\ref{Main_Eq2}) coincides with Eq.\ (\ref{Main_Eq}).
Specifing appropriate initial and boundary conditions, Eqs.\ (\ref{Main_Eq2}) 
could be solved numerically for example by the discretization scheme or 
with the Runge-Kutta method. 

For numerical solutions of the fractional 
sine-Gordon equation (\ref{D10b}) with $0 < \alpha < 2$
\be \label{fsg}
\frac{\partial^2}{\partial t^2}u(x,t) - \bar{J}_0 \; 
a_{\alpha} \frac{\partial^{\alpha}}{\partial |x|^{\alpha}} u(x,t) +
J_1 \; u(x,t) + J_2 \; \sin \left( u(x,t) \right) = 0, 
\ee
we have used two methods which are described as the following.

\begin{figure}
\centering
\rotatebox{0}{\includegraphics[width=7.8 cm,height=7.8 cm]{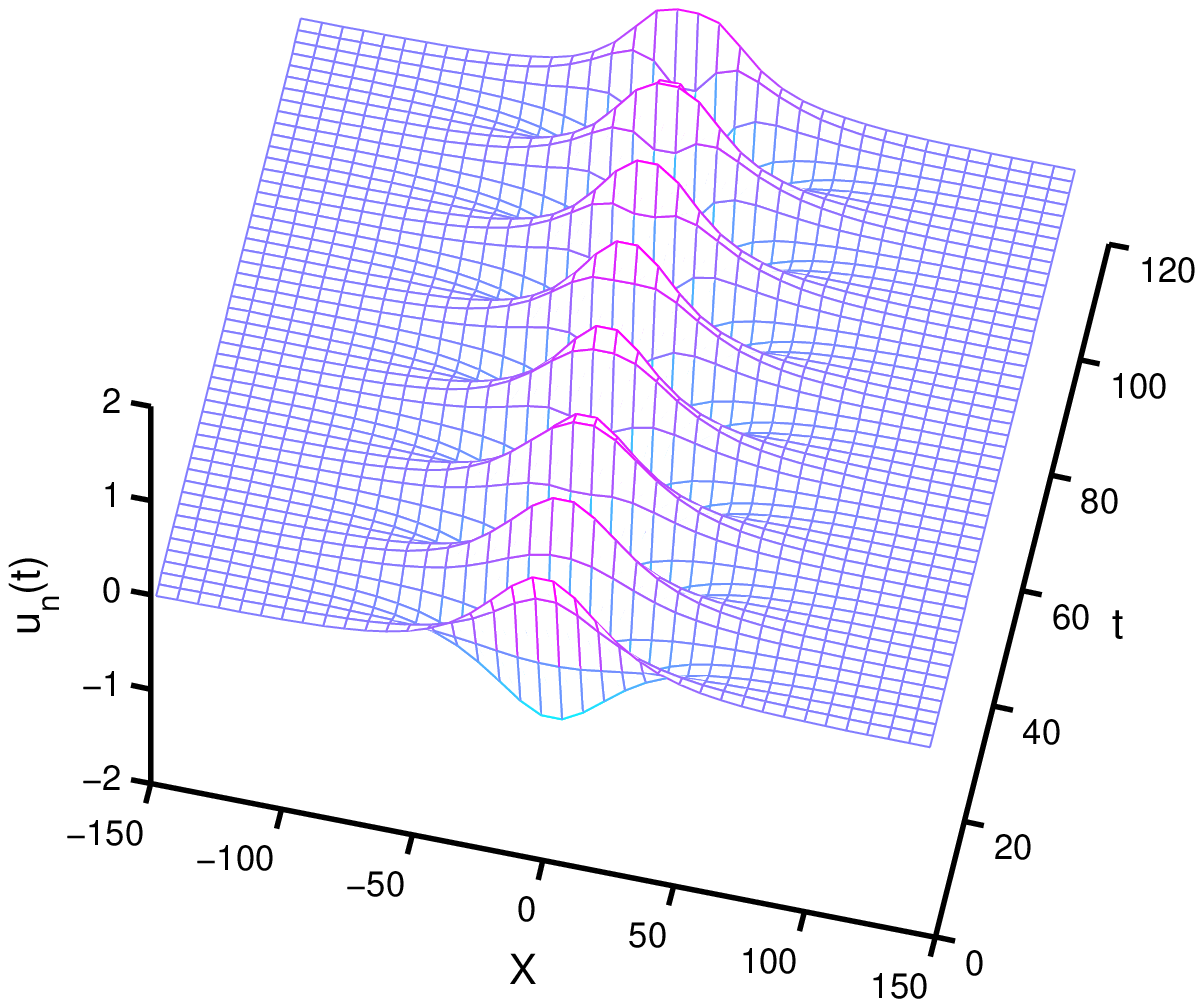}}
\rotatebox{0}{\includegraphics[width=7.8 cm,height=7.8 cm]{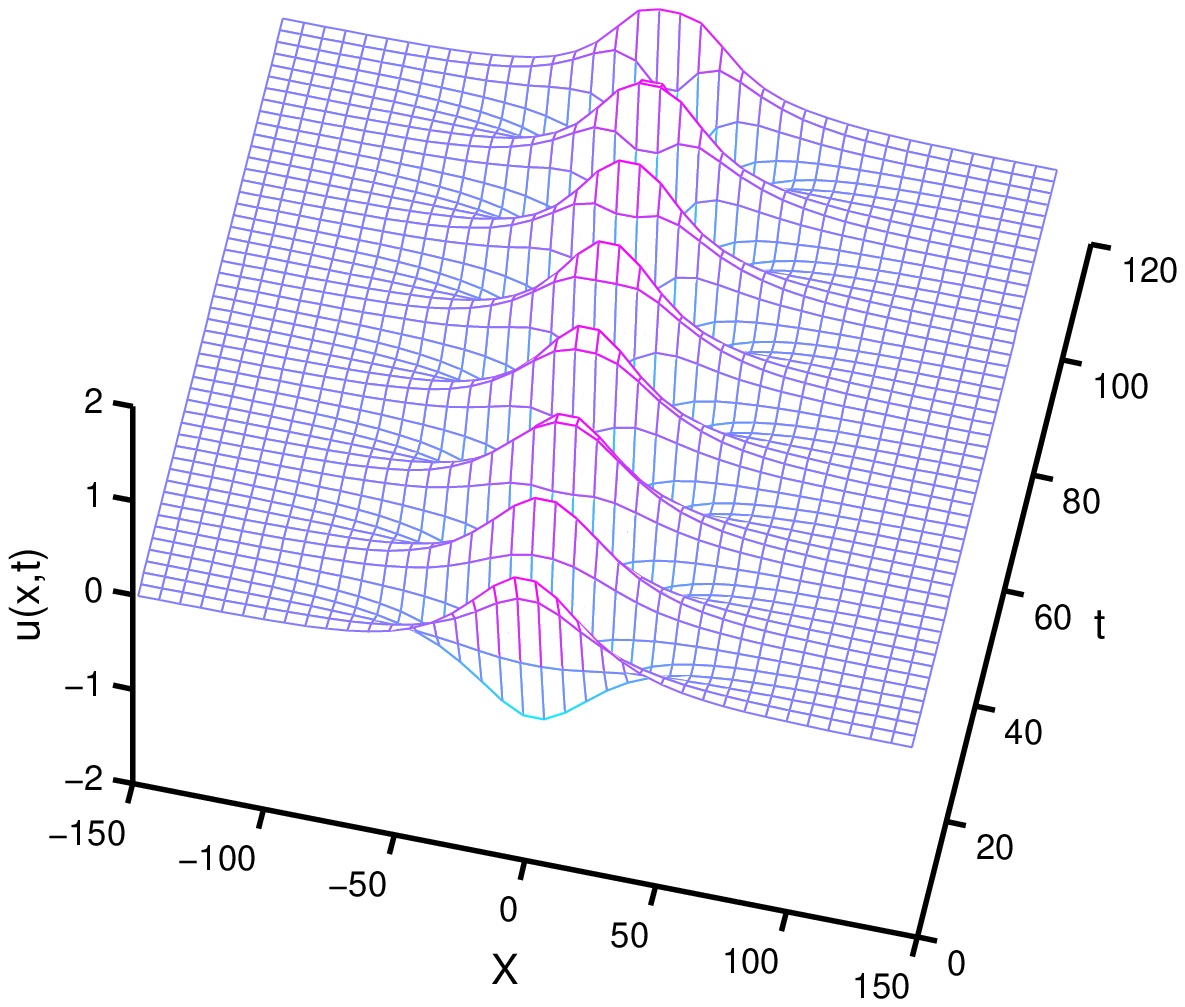}}
\rotatebox{0}{\includegraphics[width=7.8 cm,height=7.8 cm]{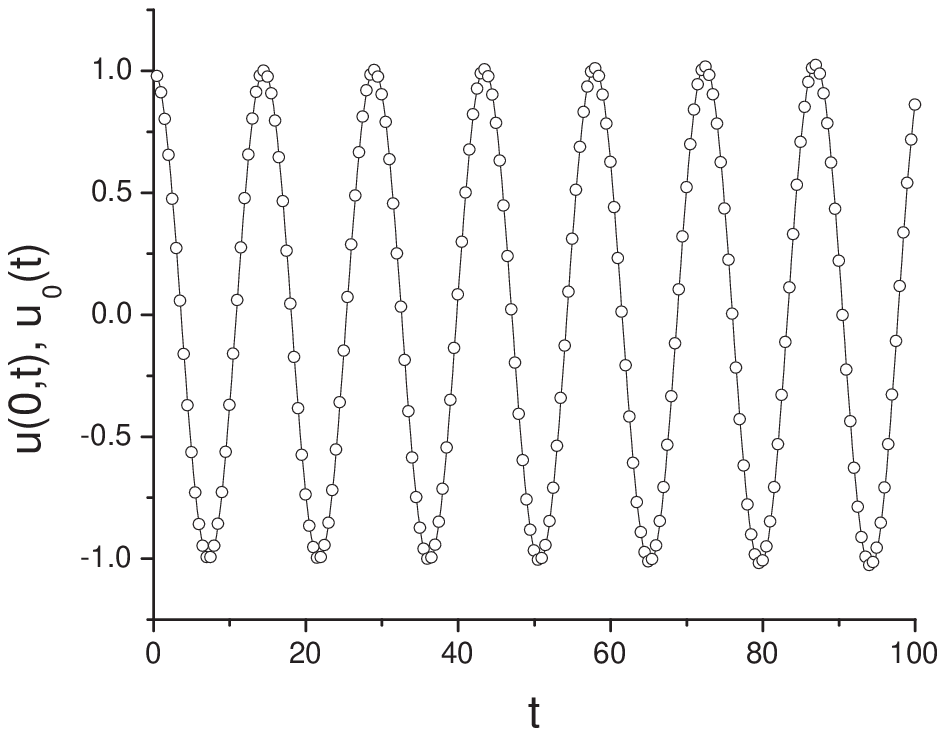}}
\caption{\label{fig2} 
Left panel: Breather-like structures obtained from the evolution 
of the system of coupled oscillators. 
Right panel: solution of the FSG equation. 
Bottom figure: Comparison of the time evolution of the central 
oscillator $u_0(t)$ (solid curve) with  the time evolution of the $u(0,t)$ (circles) obtained 
from the solution of the FSG equation. 
Exponent $\alpha=1.21$. 
}
\end{figure}

Method (a): The Riesz derivative could be represented as \cite{GM99}
\be \label{N1}
\frac{\partial^{\alpha}}{\partial |x|^{\alpha}} u(x,t)=
-\frac{1}{2 \cos(\pi \alpha /2)} 
\left({\cal D}^{\alpha}_{+}u(x,t) +{\cal D}^{\alpha}_{-} u(x,t)\right),
\ee
where ${\cal D}^{\alpha}_{\pm}$ are Riemann-Liouville 
left and right fractional derivatives defined by \cite{SKM,OS,MR,Podlubny}
\[
{\cal D}^{\alpha}_{+}u(x,t)=
\frac{1}{\Gamma(m-\alpha)} \frac{\partial^m}{\partial x^m}
\int^{x}_{-\infty} \frac{u(\xi,t) d\xi}{(x-\xi)^{\alpha-m+1}},
\]
\be \label{N2}
{\cal D}^{\alpha}_{-}u(x,t)=
\frac{(-1)^m}{\Gamma(m-\alpha)} \frac{\partial^m}{\partial x^m}
\int^{\infty}_x \frac{Z(\xi,t) d\xi}{(\xi-x)^{\alpha-m+1}},
\ee
where $m-1 < \alpha < m$. Since we seek for a numerical 
solution on a finite interval, we will also use the Riemann-Liouville left and right 
fractional derivatives defined on a finite interval $(-L,L)$ 
\[
{}_{-L} {\cal D}^{\alpha}_{x}u(x,t)=
\frac{1}{\Gamma(m-\alpha)} \frac{\partial^m}{\partial x^m}
\int^{x}_{-L} \frac{u(\xi,t) d\xi}{(x-\xi)^{\alpha-m+1}},
\]
\be \label{N3}
{}_x {\cal D}^{\alpha}_{L}u(x,t)=
\frac{(-1)^m}{\Gamma(m-\alpha)} \frac{\partial^m}{\partial x^m}
\int^{L}_x \frac{u(\xi,t) d\xi}{(\xi-x)^{\alpha-m+1}}.
\ee

\begin{figure}
\centering
\rotatebox{0}{\includegraphics[width=7.8 cm,height=7.8 cm]{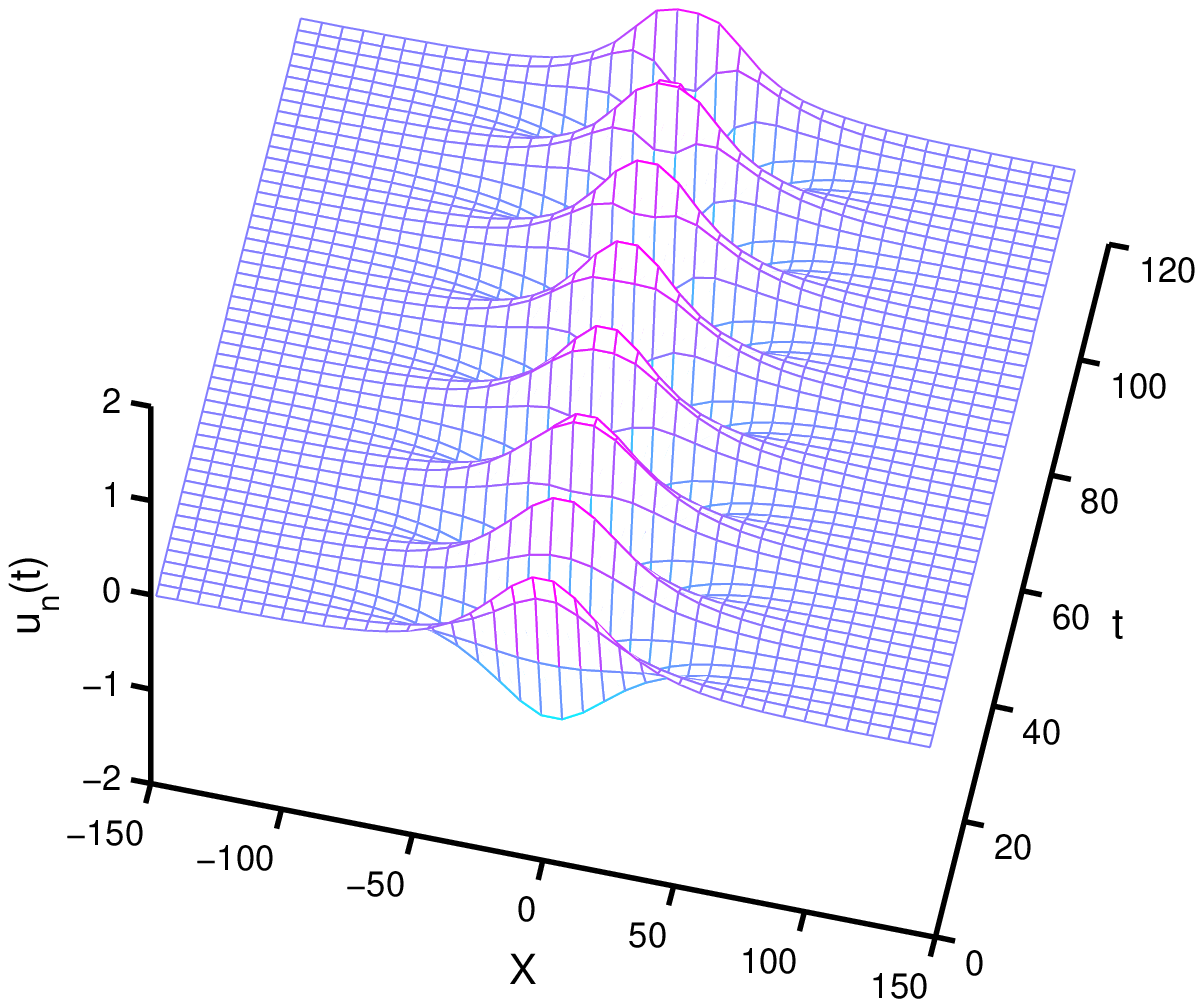}}
\rotatebox{0}{\includegraphics[width=7.8 cm,height=7.8 cm]{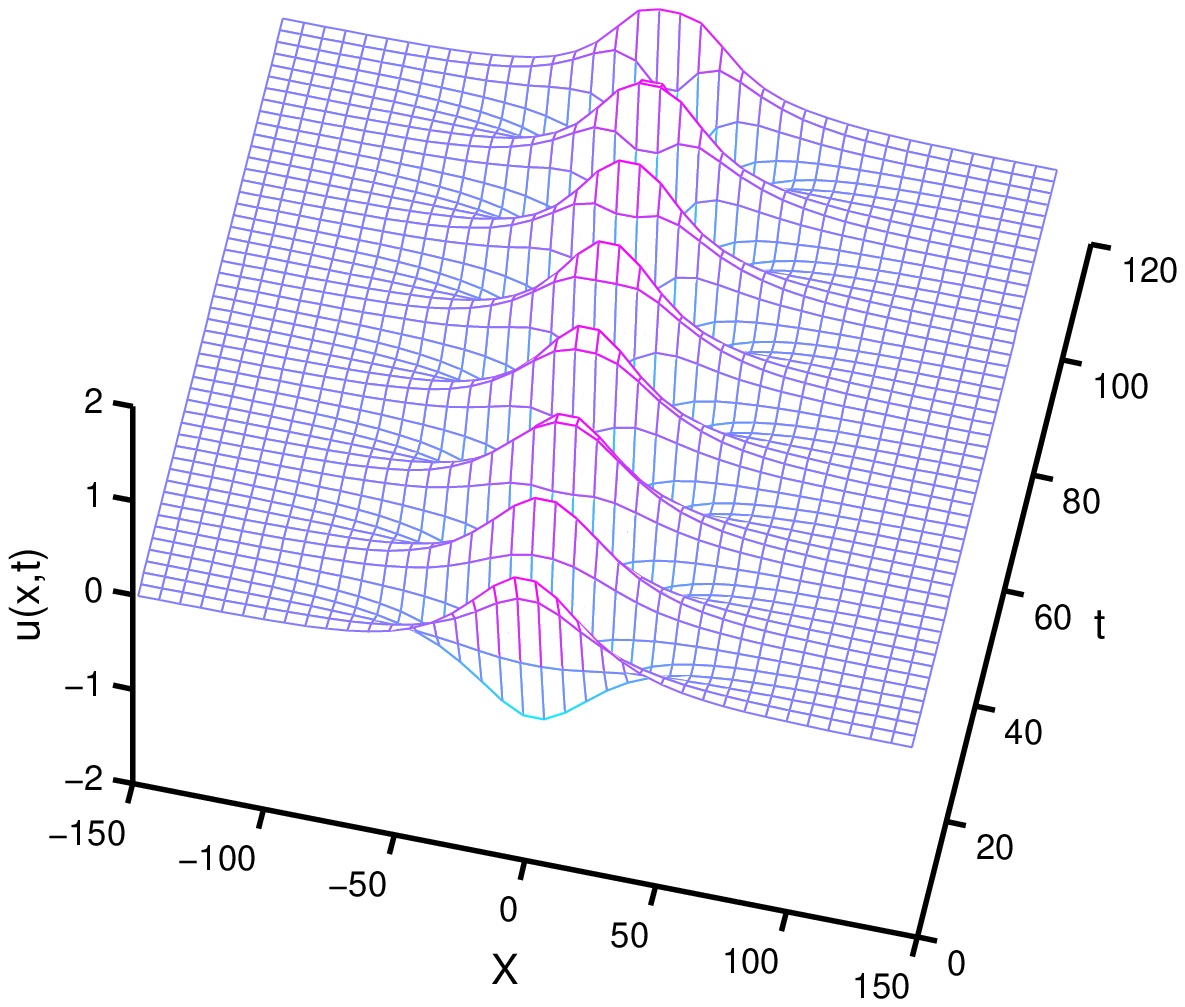}}
\rotatebox{0}{\includegraphics[width=7.8 cm,height=7.8 cm]{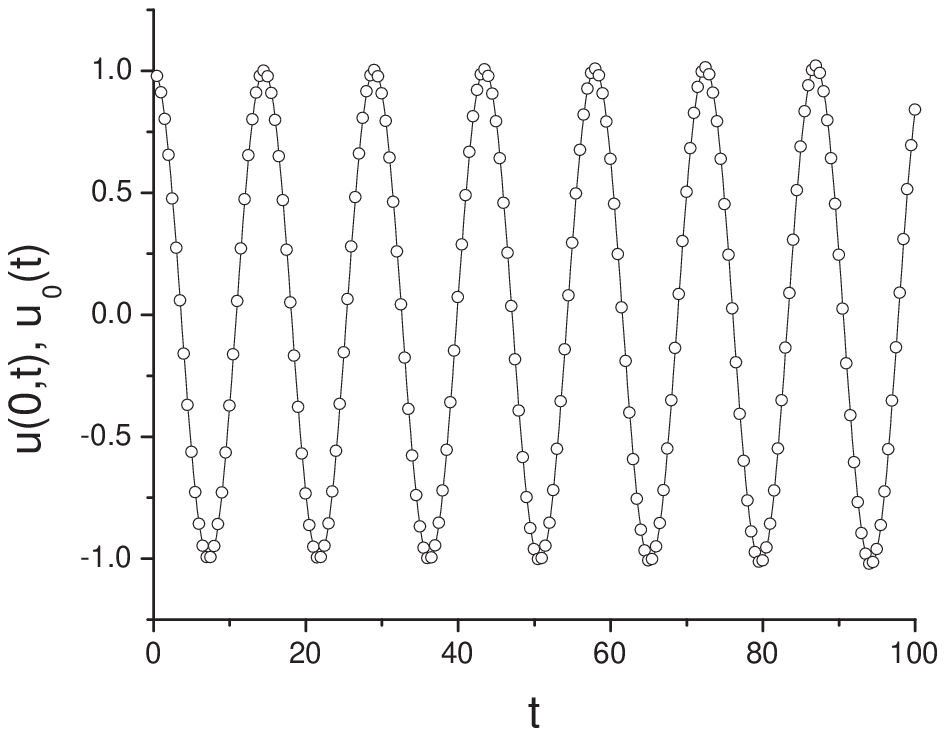}}
\caption{\label{fig2a} 
The same as in Fig.\ \ref{fig1}, but for $\alpha=1.51$.
}
\end{figure}

As the next step we approximate the Riesz fractional 
derivative as follows
\be \label{N4}
\frac{\partial^{\alpha}}{\partial |x|^{\alpha}} u(x,t) \simeq
-\frac{1}{2 \cos(\pi \alpha /2)} 
\left( {}_{-L}{\cal D}^{\alpha}_{x}u(x,t) + {}_x {\cal D}^{\alpha}_{L} u(x,t)\right).
\ee
To compute Riemann-Liouville fractional 
derivatives we have used Grunwald-Letnikov discretization scheme \cite{SKM}
\[
{}_{-L} {\cal D}^{\alpha}_{x}u(x,t)= \frac{1}{h^{\alpha}} \sum^{K_{-}}_{q=0} w_k \; u(x-qh,t),
\]
\be \label{N5}
{}_x {\cal D}^{\alpha}_{L}u(x,t)= \frac{1}{h^{\alpha}} \sum^{K_{+}}_{q=0} w_k \; u(x+qh,t),
\ee
where $h$ is the discretization parameter, 
$K_{-} =\left[(x+L)/h\right]$,  $K_{+} =\left[(L-x)/h\right]$, $[\nu]$ means integer part of $\nu$, and
\be \label{N6}
w_k = \frac{\Gamma(q-\alpha)}{\Gamma(1+q) \; \Gamma(-\alpha)}.
\ee
However, for numerical calculations it is more convenient 
to use, instead of Eq.\ (\ref{N6}), the following recursion relation
\be \label{N7}
w_k = \left(1 - \frac{\alpha+1}{q}\right) \; w_{k-1}, \; \; w_0 =1.
\ee

The numerical scheme defined by the Eqs.\ (\ref{N5}) 
could be unstable. Therefore, to obtain a stable numerical scheme, 
one can use shifted Grunwald-Letnikov formulas proposed in \cite{MT}
\be \label{N5a}
{}_{-L} {\cal D}^{\alpha}_{x} u(x,t)= \frac{1}{h^{\alpha}} \sum^{K_{-}}_{q=0} w_k \; u(x-(q-1)h,t).
\ee
\[
{}_{x} {\cal D}^{\alpha}_{L} u(x,t)= \frac{1}{h^{\alpha}} \sum^{K_{+}}_{q=0} w_k \; u(x+(q-1)h,t).
\]
In our numerical simulations both numerical methods reproduce same results. 
To ensure the stability of the integration scheme 
the analog to the Courant-Friedrichs-Lewy stability condition should be fulfiled
\be \label{N8}
\Delta t/(\Delta x)^{\alpha} < 1/2.
\ee

After substitution of Eqs.\ (\ref{N4}), (\ref{N5}), (\ref{N6}) into Eq.\ (\ref{fsg}), 
we have solved them using the explicit finite differences method for the second order time derivative 
\be 
\frac{\partial^2}{\partial t^2}u(x,t) = \frac{u(x,t+\Delta t) + u(x,t-\Delta t) - 2 u(x,t)}{(\Delta t)^2}. 
\ee
We have solved Eqs.\ (\ref{fsg}) also by the forth order Runge-Kutta method in time. 
Results of both methods appear to be the same.

Method (b): The second method for numerical solution 
of the FSG equation is based on the fact that the 
Riesz space-fractional derivative admits the explicit representation in the form of an integral \cite{SKM} with the limits of integration from zero to infinity. To use this definition for the solution of the equation on the finite interval we cut off the upper integration limit and approximate the Riesz derivative in the following way  
\be
\frac{\partial^{\alpha}}{\partial |x|^{\alpha}} u(x,t) \simeq 
\frac{\Gamma (1+\alpha)}{\pi} \sin \left( \frac{\alpha \pi}{2} \right) \int_{0}^{L} 
\frac{u(x+\eta,t)-2u(x,t)+u(x-\eta,t)}{\eta^{1+\alpha}} \; d\eta.
\ee


\section{Numerical results}

\begin{figure}
\centering
\rotatebox{0}{\includegraphics[width=7.8 cm,height=7.8 cm]{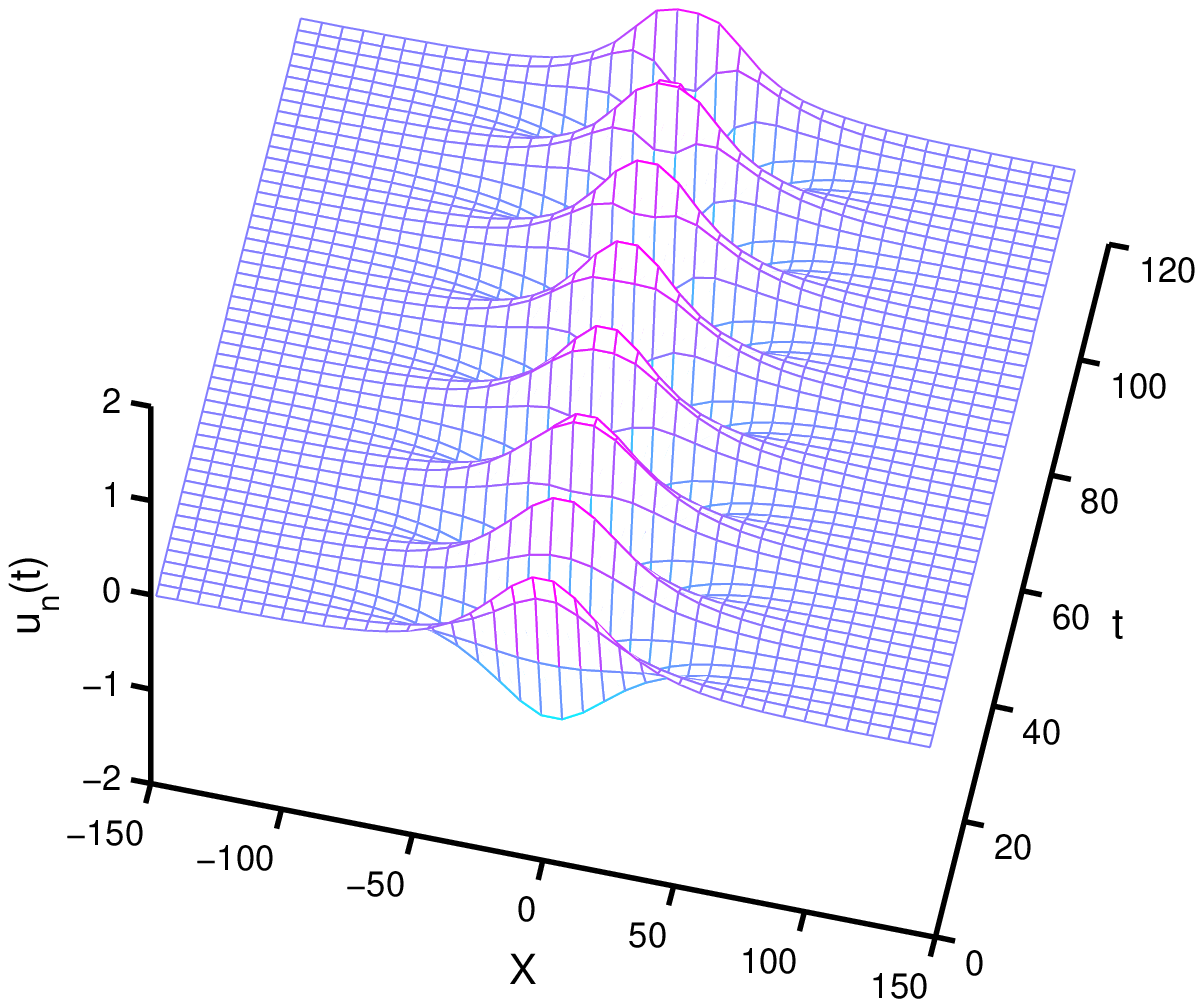}}
\rotatebox{0}{\includegraphics[width=7.8 cm,height=7.8 cm]{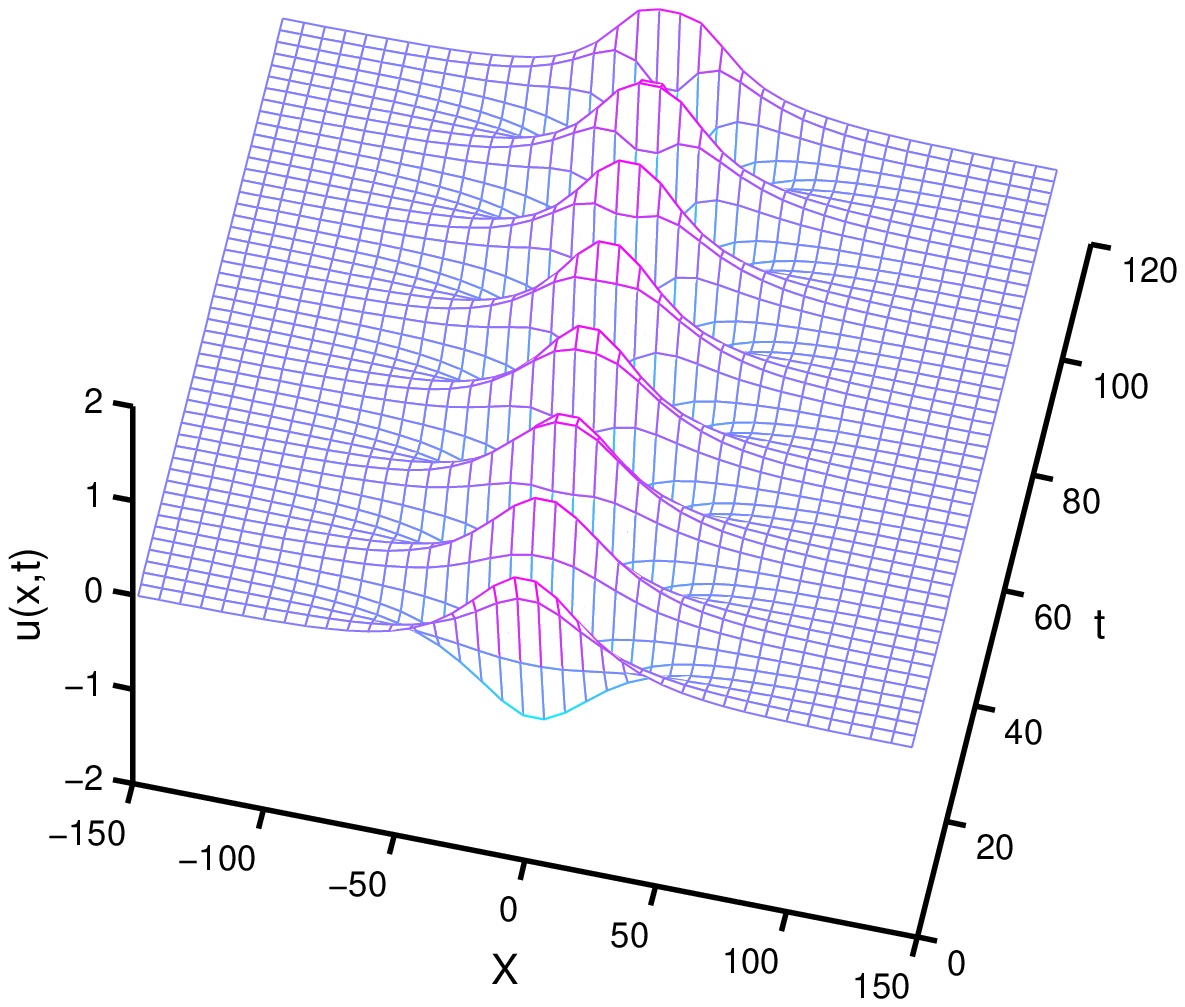}}
\rotatebox{0}{\includegraphics[width=7.8 cm,height=7.8 cm]{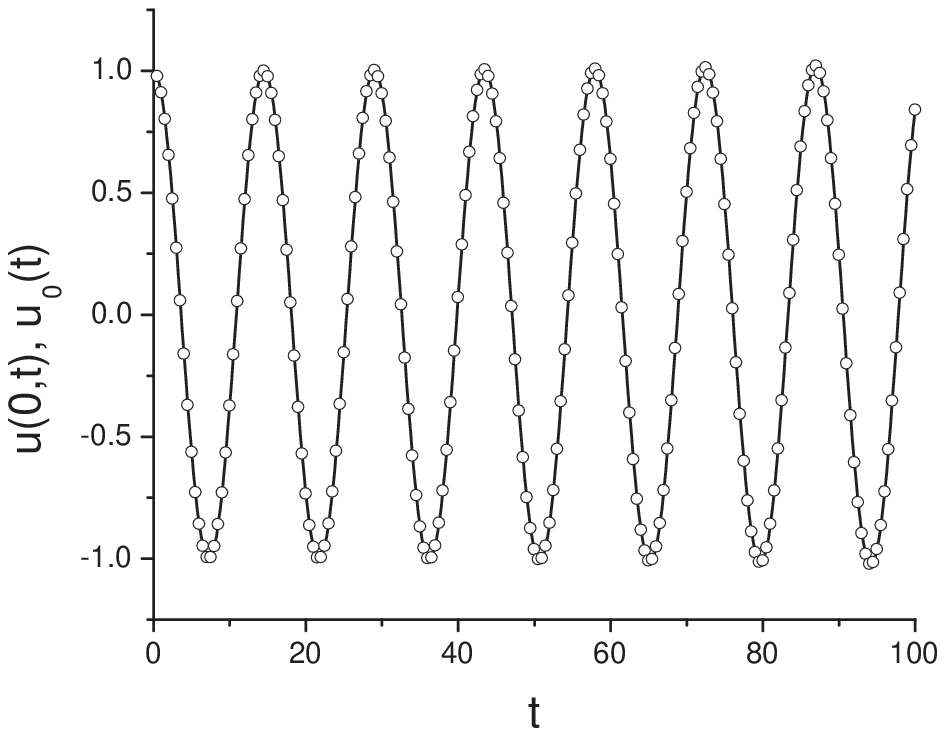}}
\caption{\label{fig2b} 
The same as in Fig.\ \ref{fig1}, but for $\alpha=1.91$.
}
\end{figure}

In this section we compare numerical solutions obtained by two different ways: solution of equations 
of motion for the system of coupled oscillators starting 
from particular initial conditions and solution of FSG with the same initial conditions. 

As the first kind of initial conditions we consider the topological soliton 
(kink) solution of the standard sine-Gordon equation
\be \label{ss}
u_n(0) = 4 \arctan \left[ \kappa \exp (x) \right],
\ee
where for the system of coupled oscillators the variable $x$ defines 
positions of oscillators, $x=n \Delta x$. The parameter $\kappa$ was fixed to $\kappa=0.001$. 
Three other constants in Eqs. (\ref{Main_Eq2}) and (\ref{fsg}) 
were $J_1 = 0.2, \; J_0 = J_2 = 0.01$. 
The long-range interaction exponent is $\alpha=1.21$. 
We consider $N+1=1001$ equally 
spaced oscillators with the distance between them $\Delta x=2L/(N+1)$, $L=500$ 
and boundary conditions $u_{n+2L+1}(t) = u_n(t) + 2\pi$. 
For the solution of FSG equation the discretization parameter $h$ was choosen to be $\Delta x/2$.  

The time evolution of this 
initial function is shown in Fig.\ \ref{fig1}. The left panel 
represents the solution of the equations of motion for the discrete 
system of coupled oscillators, while the solution of the FSG equation 
is shown in the right panel. As it is seen from 
the figure, solutions of both systems are very similar to each other. 
On the bottom panel we compare the behaviour of $u_0(t)$ (solid line) 
and $u(0,t)$ (circles). The fractional order of the space derivative 
is reflected in the time oscillations of the initial soliton-like 
profile which is the exact stationary solution for the standard 
sine-Gordon equation with $\alpha=2$.

\begin{figure}
\centering
\vspace{-1.5cm}
\includegraphics[width=8 cm,height=6.5 cm]{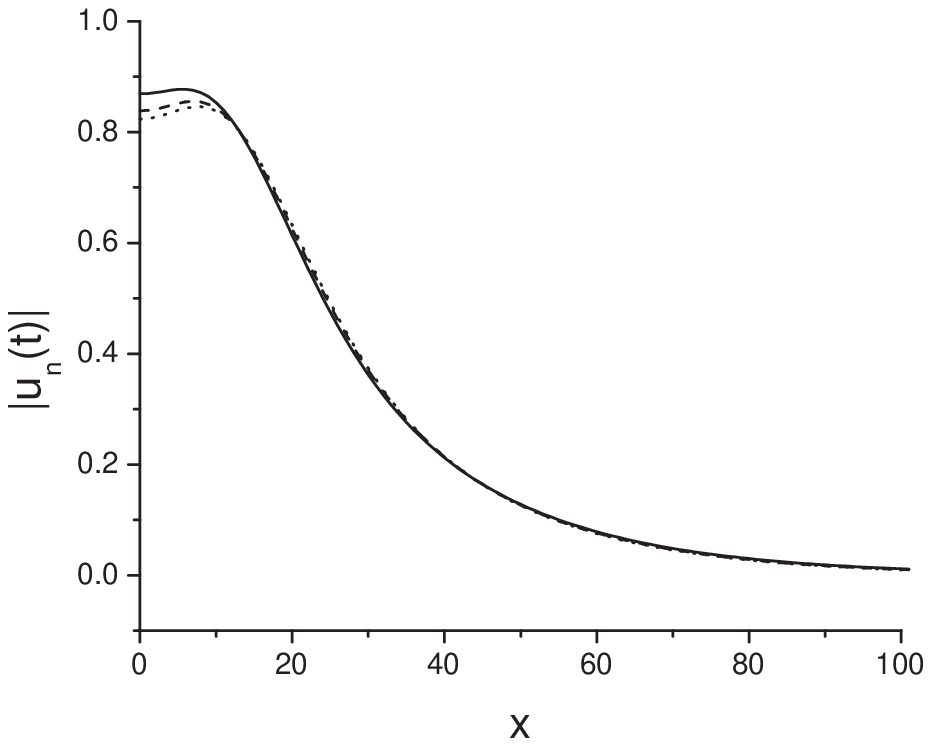}
\includegraphics[width=8 cm,height=6.5 cm]{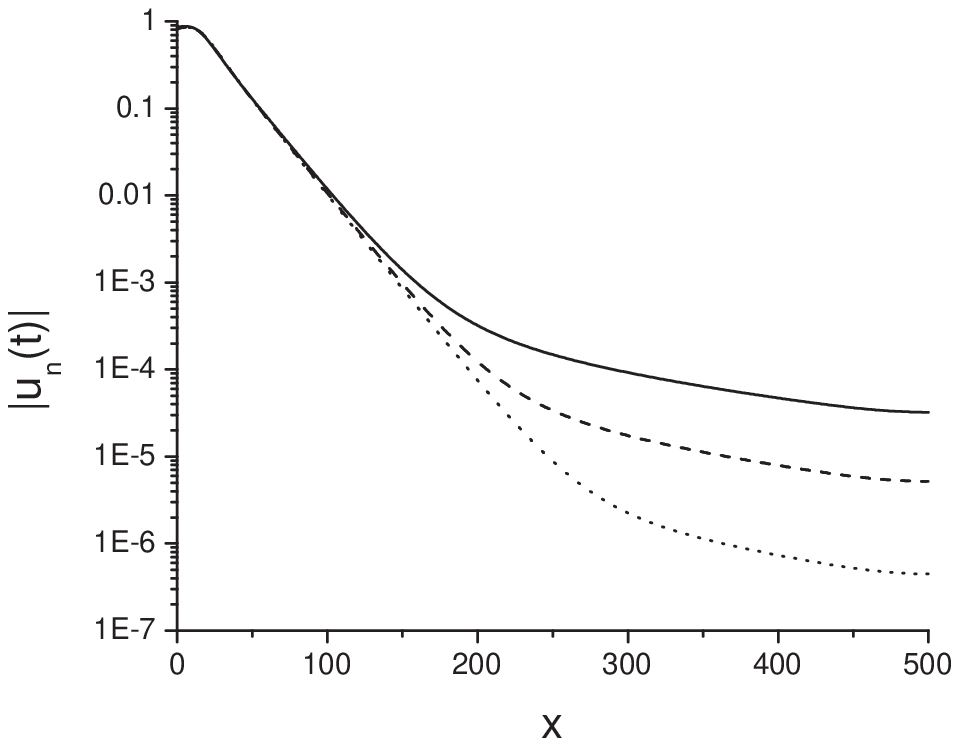}
\includegraphics[width=8 cm,height=6.5 cm]{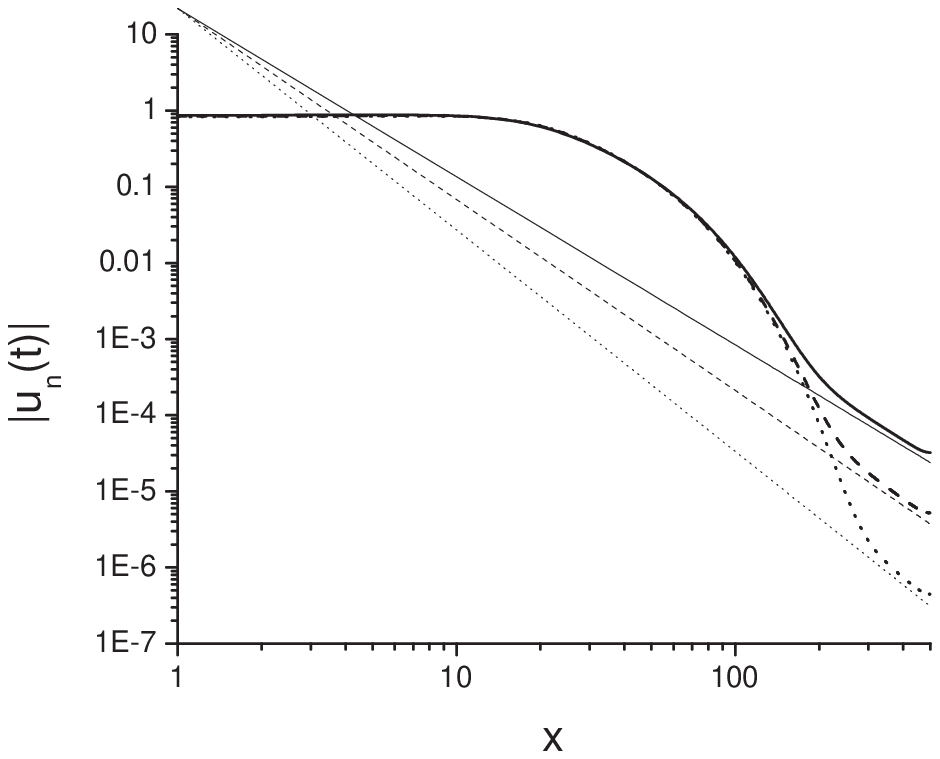}
\caption{\label{fig3} 
Profiles of the breather-like solutions obtained for the system of coupled 
oscillators for $t=100$ and $\alpha=1.21, 1.51, 1.91$ (solid, dashed and dotted curves respectively) 
in normal, the semi-logarithmic and in the double-logarithmic scales. Straight lines correspod to $f(x) \sim 1/x^{1+\alpha}$.
}
\end{figure}

As a second kind of profile we consider two-parametric standing soliton 
solution of the standard sine-Gordon equation
\be \label{N0}
u_n(0) = 4 \arctan \left[ \frac{\nu}{\kappa \cosh (x)} \right],
\ee
where $\nu=1$ and $\kappa=0.05$. The constant was choosen such that the ratio wave length 
to the distance between oscillators is small. 
Three other constants in Eqs.\ (\ref{Main_Eq2}), (\ref{fsg}) were fixed to 
$J_0 = 0.01, \; J_1=J_2 = 0.1$. The number of oscillators 
$N+1=1001$, the distance between them $\Delta x=2L/(N+1)$ with $L=500$, 
and the periodic boundary conditions were applied $u_{n+2L+1}(t) = u_n(t)$.
For the solution of the FSG equation the discretization we choose $h=\Delta x/2$. 

Solutions of both equations appear in the form 
of a breather-like structures. They are ploted 
for different values of $\alpha$ in Figs.\ \ref{fig2}, \ref{fig2a} and \ref{fig2b}. 
Solutions for the system of 
coupled oscillators are presented on the left panels of 
Fig.\ \ref{fig2}, \ref{fig2a} and \ref{fig2b}, while solutions 
of the fractional sine-Gordon equation Eq.\ (\ref{fsg}) are ploted 
on the right panels. 

From these figures we conclude that 
in the infrared limit solutions of both systems are almost identical.
Another conclution, that one can draw, is the low sensitivity of the 
core breather-like structure on the exponent $0< \alpha < 2$. 
In Fig.\ \ref{fig3} the snapshots of the breather-like solutions obtained 
for the same initial conditions are plotted for the time instant $t=100$. 
The long-range interaction mostly contributs to the behaviour of the tails 
of the breather-like solutions. The exponential spatial decay 
for short distances is modified by the power-law for large distances 
\cite{Br4, GF} due to non-local interaction $\sim 1/x^{1+\alpha}$. 
In Fig.\ \ref{fig3} (middle and low panels) the profiles of the breather-like solutions 
are ploted in semi-logarithmic and in double-logarithmic scales. 
The crossover from the exponential decay of the 
profile for a short distances to the power-law decay for a large distances is explicitly seen. 
With the increasing value of the parameter $\alpha$ the cross-over is shifted 
for longer time with the agreement to the reduction of the power-law 
interaction to the nearest-neighbor one for $\alpha \rightarrow \infty$. 
With a decrease of $\alpha$ the exponential part of the decay 
shrinks while the power-law decay is getting broad. 

\section{Conclusion}

One-dimensional chain of classical interacting oscillators 
serve as a model for numerous applications 
in physics, chemistry, biology, etc. Long-range interactions, i.e., with
forces proportional to $1/|x|^{1+\alpha}$ are important type of interaction for complex media. 
An interesting situation arrises when, for some reasons, the power $\alpha$ is non-integer. 
A remarkable feature of these interactions is the existence
of a transform operator that replaces the set of 
coupled individual oscillator equations 
into the continuous medium equation
with the space derivative of 
order $\alpha$, where $0<\alpha<2$, $\alpha\not=1$.
Such transformation is an approximation and 
it appears in the infrared limit.
This limit helps to consider different models and related phenomena
in unified way applying
different tools of fractional calculus.

Periodic space-localized oscillations 
which arise in discrete and continuous nonlinear systems have been widely studied in systems 
with short-range interactions. Here, the system with long-range interactions was considered. 
The method to map the set of equations of motion  
onto the continuos fractional order differential equation is developed 
in terms of the transform operator. It is known that the properties 
of a system with long-range interactions are very different 
from short-range one. The method offers a new tool for the analysis 
of different soliton- and breather-type solutions in such systems. 
 
Vice versa, we hope that in some cases the obtained transform operator can be 
used to improve methods for numerical solutions of equations with fractional derivatives. 

In quantum case the application of the transform operator would be highly interesting for example to a discrete nonlinear Schr{\'o}dinger-like equations with the power-law interactions.  

\section*{Acknowledgments}

This work was supported by the Office of Naval Research,
Grant No. N00014-02-1-0056, the U.S. Department
of Energy Grant No. DE-FG02-92ER54184, and the NSF
Grant No. DMS-0417800. 
V.E.T. thanks the Courant Institute of Mathematical Sciences
for support and kind hospitality. N.K. wish to thank A. Gorbach for discussions.

\end{document}